\documentclass[showpacs,
aps,prd,
,nofootinbib,
floats]{revtex4}
\usepackage{graphicx}
\usepackage{amssymb,latexsym}
\usepackage[draft=false]{hyperref}

\def\lsim{\hbox{ \raise.35ex\rlap{$<$}\lower.6ex\hbox{$\sim$}\ }}
\def\gsim{\hbox{ \raise.35ex\rlap{$>$}\lower.6ex\hbox{$\sim$}\ }}
\def\setC{\mathbb{C}}
\def\setR{\mathbb{R}}

\begin{document}

\title{Consistency of cosmic strings with cosmic microwave background
measurements}

\author{Jonathan Rocher}
\email{rocher@iap.fr}
\affiliation{Institut d'Astrophysique
de Paris, {${\cal G}\setR\varepsilon\setC{\cal O}$}, FRE 2435-CNRS, 98bis
boulevard Arago, 75014 Paris, France.}

\author{Mairi Sakellariadou}
\email{msakel@cc.uoa.gr, sakella@iap.fr}
\affiliation{Division of Astrophysics, Astronomy, and Mechanics, Department
of Physics,  University
of Athens, Panepistimiopolis, GR-15784 Zografos, Hellas, and
\\
Institut d'Astrophysique
de Paris, 98bis boulevard Arago, 75014 Paris, France.}


\begin{abstract}
In the context of SUSY GUTs, GUT scale cosmic strings formed at the
end of hybrid inflation are compatible with currently available CMB
measurements. The maximum allowed cosmic strings contribution to the
CMB data constrains the free parameters (mass scales, couplings) of
inflationary models. For F-term inflation either the superpotential
coupling must be fine tuned or one has to invoke the curvaton mechanism.
D-term inflation should be addressed in the framework of SUGRA. We find that
the cosmic strings contribution is not  constant, contrary to some
previous results. Using CMB measurements we constrain the gauge and
the superpotential couplings.
\end{abstract}

\maketitle

Topological defects are the natural outcome of phase transitions
accompanied by Spontaneously Symmetry Breaking (SSB) as the Universe
expands.  Among the various types of defects only Cosmic Strings (CS)
are not dangerous to overclose the Universe. To get rid of the
undesired defects, monopoles and domain walls, one usually employs the
mechanism of inflation. Starting from a large gauge group in the
context of Supersymmetric (SUSY) Grand Unified Theories (GUTs), in
Ref.~\cite{rjs} there were considered all possible SSB schemes down
to the Standard Model (SM) gauge group. Placing one or more inflationary
eras to dilute the harmful stable defects, it was found that CS are
always left behind.  From the observational point of view however, strong
constraints~\cite{bprs,p} are placed on the allowed CS contribution to
the angular power spectrum of the Cosmic Microwave Background (CMB)
temperature anisotropies. This finding led some colleagues to
conclude that CS, at least of the GUT scale, are ruled out, since
their theoretical predictions seemed to be inconsistent with
measurements. The aim of our study is to show that this is by no means
the case.  We find instead that the maximum allowed CS contribution to
the CMB measurements places upper limits on the inflationary scale
(which is also the CS energy scale), or equivalently on the coupling
constant of the superpotential.  Our analysis, in the framework of
supersymmetric hybrid inflation, is given below first for F- and then
for D-term type, where we also consider Supergravity (SUGRA)
corrections, which turn out to be essential.

\subsection{F-term inflation}
F-term inflation is based on the supersymmetric renormalisable superpotential
\begin{equation}\label{superpot}
W_{\rm infl}^{\rm F}=\kappa S(\Phi_+\Phi_- - M^2)~,
\end{equation}
where $S, \Phi_+, \Phi_-$ are three chiral superfields, and $\kappa$,
$M$ are two constants.  We denote hereafter the superfield and its
scalar component by $S$. Taking into account the one-loop radiative
corrections to the scalar potential along the inflationary valley, the
effective potential reads~\cite{DvaShaScha,Lazarides,SenoSha,rs}
\begin{equation}
\label{VexactF}
V_{\rm eff}^{\rm F}(|S|)=\kappa^2M^4\left\{1+\frac{\kappa^2
\cal{N}}{32\pi^2}\left[2\ln\frac{|S|^2\kappa^2}{\Lambda^2}+
(z+1)^2\ln(1+z^{-1})+(z-1)^2\ln(1-z^{-1})\right]\right\}~,
\end{equation}
where $z=|S|^2/M^2\equiv x^2$, $\Lambda$ is a renormalisation scale
and $\cal{N}$ stands for the dimensionality
of the representation to which the complex scalar components $\phi_+,
\phi_-$ of the chiral superfields $\Phi_+, \Phi_-$ belong.

On large angular scales, the main contribution to the CMB anisotropies
is given by the Sachs-Wolfe effect. The quadrupole anisotropy has a
contribution from the quantum fluctuations of the inflaton field and,
in our model, a contribution from the CS network. The inflaton
contribution (splitted into scalar and  tensor parts) can be computed
analytically, while the CS contribution is only derived from numerical
simulations~\cite{landriau}. Their sum reads~\cite{rs},
\begin{equation}\label{eqnumF}
\left({\delta T\over T}\right)_{\rm Q-tot}^2 \sim
y_{\rm Q}^{-4}\left({\kappa^2 \mathcal{N}\,
N_{\rm Q}\over 32\pi^2}\right)^2
\left[\frac{64N_{\rm Q}}{45\cal N} x_{\rm Q}^{-2}y_{\rm Q}^{-2}f^{-2}(x_Q^2)
 +\left(\frac{0.77 \kappa}{\pi}\right)^2 + 324\right]~,
\end{equation}
where $x_{\rm Q}^2=|S_{\rm Q}|^2/M^2$ (the index Q denotes the scale
corresponding to the quadrupole anisotropy), $y_{\rm Q}^2$ is defined by
\begin{equation}
\label{yQ}
y_{\rm Q}^2=\int_1^{x_{\rm Q}^2}\frac{{\rm d}z}{zf(z)}~,
\end{equation}
and the number of e-foldings $N_{\rm Q}$ is
\begin{equation}
N_{\rm Q}=\frac{4\pi^2}{\kappa^2\cal N}\frac{M^2}{M_{\rm Pl}^2}\,y_{\rm Q}^2~,
\end{equation}
 with $M_{\rm Pl}$ the reduced Planck mass [$M_{\rm Pl}= {(8\pi
G)^{-1/2}}$], and
$f(z)=(z+1)\ln(1+z^{-1})+(z-1)\ln(1-z^{-1})$.  The {\sl l.h.s.} of
Eq.~(\ref{eqnumF}) is normalised to the COBE data. For given $\kappa,
N_{\rm Q}$, and $\mathcal{N}$, Eq.~(\ref{eqnumF}) can be solved
numerically~\cite{rs} to get $x_{\rm Q}$, which is the only unknown
quantity. Thus, we  are able to compute $y_{\rm Q}$ and $M$. We
find~\cite{rs} that the  mass scale $M$ grows very slowly with
$\cal N$ and it is of the order of $10^{15}$ GeV. Hereafter we use
$\mathcal{N}=\mathbf{3}$ since this is the most generic
value in SUSY GUTs~\cite{rs}.  $M$ gives the scale of inflation as well as the
mass scale of cosmic strings formed at the end of the inflationary
era. The mass scale $M$ is related to the coupling $\kappa$
through
\begin{equation}\label{Mdekappa}
\frac{M}{M_{\rm Pl}}=\frac{\sqrt{N_{\rm Q} \cal N}\,\kappa}{2\,
\pi\,y_{\rm Q}}~.
\end{equation}

\begin{figure}[htbp]
\includegraphics[scale=.4]{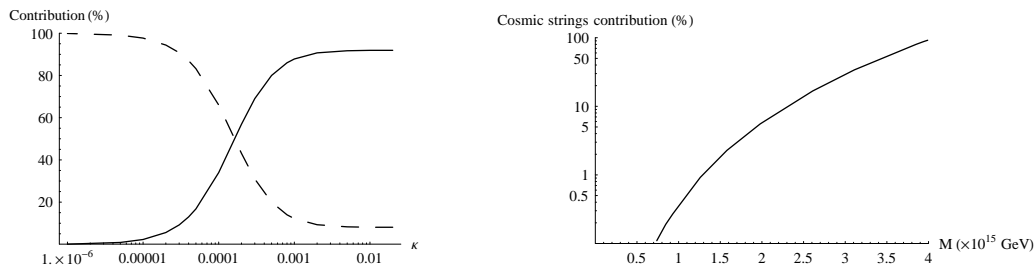}
\caption{On the left, cosmic strings (full line) and inflaton field
(broken line) contributions to the CMB fluctuations as a function of
the coupling $\kappa$.  On the right, the cosmic strings contribution
as a function of the mass scale of inflation $M$.}\label{PRLresult}
\end{figure}

The CS contribution to the CMB, denoted by ${\cal A}_{\rm CS}$, is a
function of the coupling $\kappa$, or equivalently of the  mass
scale $M$.  In Fig.~\ref{PRLresult} we show the inflaton and cosmic
strings contributions with respect to $\kappa$ and $M$.

We find that ${\cal A}_{\rm CS}$ is consistent
with the WMAP CMB data, namely \cite{p} ${\cal A}_{\rm CS}\lsim 9\%$, provided
\begin{equation}
M\lsim 2.2\times 10^{15} {\rm GeV} ~~\Leftrightarrow ~~\kappa \lsim
3\times10^{-5}~.
\end{equation}

Our finding is important: CS of the GUT scale are
allowed. On the other hand, for this statement to be true,
hybrid SUSY inflation losses some of its appeal
since it is required some amount of fine tuning of its free parameter,
namely $\kappa$
should be of the order of $10^{-5}$ or smaller.
This constraint on $\kappa$ is in agreement with the one given in
Ref.~\cite{kl}.
The parameter $\kappa$ is also subject to the gravitino constraint which
imposes an upper limit to the reheating temperature, to avoid gravitino
overproduction. Within SUSY GUTs, and assuming a see-saw mechanism to
give rise to massive neutrinos, the inflaton field will
decay during reheating into pairs of right-handed neutrinos.
Using the constraints on the see-saw mechanism it is
possible~\cite{SenoSha,rs}
to convert the constraint on the reheating
temperature to a constraint on the coupling parameter $\kappa$, namely
$\kappa \lesssim 8.2\times 10^{-3}~$,
which is clearly a weaker constraint.

The superpotential coupling  $\kappa$ is allowed to get higher values, namely
it can approach the upper limit permitted by the gravitino constraint,
if one employs the curvaton mechanism~\cite{lw2002,mt2001}.
Such a mechanism can be easily
accommodated within SUSY theories, where one expects to have a number
of scalar fields. The extra curvaton contribution to the quadrupole
anisotropy reads~\cite{rs}
\begin{equation}
\left({\delta T\over T}\right)_{\rm curv}^2 =
y_Q^{-4}\,\left({\kappa^2 \mathcal{N} N_{\rm Q}\over 32\pi^2} \right)^2\,
\left[ \left(\frac{16}{81\pi\sqrt3}\right)\,
\kappa\, \left(\frac{M_{\rm Pl}}{{\cal\psi}_{\rm init}}\right)\right]^2~,
\end{equation}
where ${\cal\psi}_{\rm init}$ denotes the initial value of the
curvaton field. For fixed $\kappa$, the CS contribution decreases
rapidly as ${\cal\psi}_{\rm init}$ decreases. Thus, the WMAP
measurements lead to an upper limit on ${\cal\psi}_{\rm init}$, namely
${\cal\psi}_{\rm init}\lsim 5\times 10^{13}(\kappa/10^{-2})$ GeV.  This limit
holds for $\kappa$ in the range $[5\times 10^{-5}, 1]$; for lower values of
$\kappa$, the cosmic strings contribution is always suppressed and
thus lower than the WMAP limit.

\begin{figure}[htbp]
\includegraphics[scale=.45]{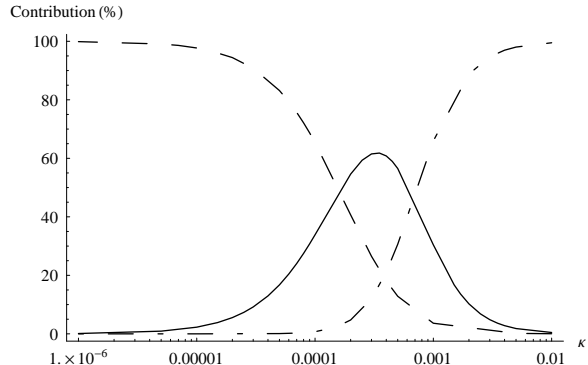}\\
\caption{Contribution from the three different sources to the CMB
anisotropies as a function of the superpotential coupling
$\kappa$, for ${\cal\psi}_{\rm init}=10^{13} {\rm GeV}$. The three
curves show the contributions from cosmic strings (full line),
inflaton (curve with broken line) and curvaton fields (curve with
lines and dots).}\label{contribPRL}
\end{figure}

Including SUGRA corrections to the F-term inflation we find~\cite{rs} no
difference in the calculated $\delta T/T$.
This is expected since the  value of the inflaton
field is several orders of magnitude below the Planck scale.

\subsection{D-term inflation}
D-term inflation is derived from the superpotential
\begin{equation}
\label{superpotD}
W^{\rm D}_{\rm infl}=\lambda S \Phi_+\Phi_-~.
\end{equation}
D-term inflation requires the existence of a nonzero Fayet-Illiopoulos
term $\xi$, permitted only if an extra U(1) symmetry is
 introduced. As previously, we calculate in the SUSY framework the
radiative corrections
 leading to the effective potential~\cite{rs}
\begin{equation}
\label{VexactD}
V^{\rm D}_{\rm eff}(|S|) =
\frac{g^2\xi^2}{2}\left\{1+\frac{g^2}{16\pi^2}
\left[2\ln\frac{|S|^2\lambda^2}{\Lambda^2}+
(z+1)^2\ln(1+z^{-1})+(z-1)^2\ln(1-z^{-1})\right]\right\}~,
\end{equation}
where $z=\lambda^2 |S|^2/(g^2\xi)\equiv x^2$, with $g$ the gauge coupling of
the U(1) symmetry and $\xi$ the Fayet-Illiopoulos term, chosen to be
positive.  We find that in the absence of the curvaton mechanism, the
quadrupole anisotropy reads~\cite{rs}
\begin{equation}\label{eqnumD}
\left({\delta T\over T}\right)_{\rm Q-tot} \sim
\left\{y_{\rm Q}^{-4}\left({\lambda^2 \,
N_{\rm Q}\over 16\pi^2}\right)^2
\left[\frac{16N_{\rm Q}}{45} x_{\rm Q}^{-2}y_{\rm Q}^{-2}f^{-2}(x_Q^2)
 +\left(\frac{0.77 g}{\sqrt{2}\pi}\right)^2 + 324\right]\right\}^{1/2}~,
\end{equation}
where the number of e-foldings is
\begin{equation}
N_{\rm Q}=\frac{2\pi^2}{\lambda^2}
\frac{\xi}{M_{\rm Pl}^2} \int_{z_{\rm end}}^{z_Q} \frac{dz}{zf(z)}~,
\end{equation}
with $z_{\rm end}=\lambda^2|S_{\rm end}|^2/(g^2\xi)$.
 Since inflation ends when either the symmetry is spontaneously broken
or the slow roll conditions are violated, assuming the gravitino constraint on
$\lambda$, we get~\cite{rs} $z_{\rm
end}\simeq 1$. The {\sl l.h.s.} of Eq.~(\ref{eqnumD}) is normalised to
the COBE data, i.e., $\left(\delta T/ T\right)_{\rm Q}^{\rm COBE} \sim
6.3\times 10^{-6}$.

We compute the mass scale of the SSB, given by $M_{\rm D}=\sqrt\xi$,
and find that the dependence of $M_{\rm D}$ on $\lambda$ is
very close to the one obtained for F-term inflation.  The cosmic strings
contribution is also very similar to the F-term case, implying
that within the SUSY framework, the same conclusions hold,
namely $\lambda\lesssim 3\times 10^{-5}$.

\begin{figure}[htbp]
\includegraphics[scale=.45]{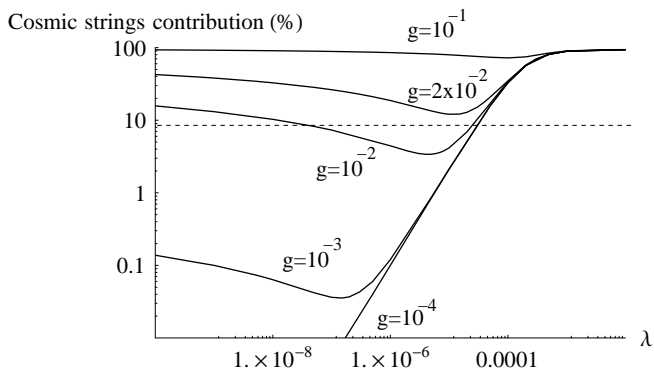}\\
\caption{Cosmic strings contribution to the CMB fluctuations as a function
of the superpotential coupling $\lambda$ for different values of the gauge
coupling $g$. The maximal contribution alowed by WMAP is represented by a
dotted line.}\label{PRL4}
\end{figure}

However, the dependence of $x_{\rm Q}$ on the superpotential coupling
$\lambda$ results to a higher value of the inflaton field $S_{\rm Q}$
 than in the F-term case, especially for large values of the gauge
coupling $g$. This implies that the correct analysis has to be done in
the framework of SUGRA.   The SUSY analysis will be a limit of the
SUGRA study for small values of the inflaton field.  Some previous
studies~\cite{rj,jap} found in the literature kept only the first term of
the radiative corrections. We find that it is necessary to perform
the analysis using the {\sl full} effective potential,
which for minimal supergravity reads~\cite{rs}
\begin{equation}
\label{VexactDsugra}
V^{\rm D-SUGRA}_{\rm eff} =
\frac{g^2\xi^2}{2}\left\{1+\frac{g^2}{16\pi^2}
\left[2\ln\frac{|S|^2\lambda^2}{\Lambda^2}\exp\left({|S|^2\over M_{\rm
Pl}^2}\right)+
(z+1)^2\ln(1+z^{-1})+(z-1)^2\ln(1-z^{-1})\right]\right\}~,
\end{equation}
where $z=[\lambda^2 |S|^2/(g^2\xi)]\exp(|S|^2/M_{\rm Pl}^2)\equiv x^2$.  The
constant term in the potential is identical to the SUSY case, but its
first derivative is modified~\cite{rs} by the factor $(1+|S|^2/M_{\rm
Pl}^2)$. The number of e-foldings is~\cite{rs}
\begin{equation}
N_{\rm Q}={2\pi^2\over g^2}\, \int_1^{x_{\rm Q}^2}
\frac{{\rm W}(c\,z)}
{z^2 f(z)[1+{\rm W}(c\,z)]^2}\, {\rm d}z~,
\end{equation}
where ${\rm W}(x)$ denotes the ``W-Lambert function'' defined by
${\rm W}(x)\exp[{\rm W}(x)]=x$, and $c\equiv
(g^2 \xi)/(\lambda^2 M_{\rm Pl}^2)$. The number of e-foldings
$N_{\rm Q}$ is thus a function of $c$
and $x_{\rm Q}$, for $g$ fixed.  Setting $N_{\rm Q}=60$ we obtain
a numerical relation between $c$ and $x_{\rm Q}$ which allows us
to construct a function $x_{\rm Q}(\xi)$ and  express the three
contributions to the CMB only as a function of $\xi$. Thus, we solved
\begin{equation}\label{eqnumDsugra}
\left({\delta T\over T}\right)_{\rm Q-tot} \sim
\frac{\xi}{M_{\rm Pl}^2}\left\{ \frac{\pi^2}{90g^2}x^{-4}_{\rm
Q}f^{-2}(x^2_{\rm Q}) \frac{{\rm W}(c\,x^2_{\rm Q})} {\left[1+{\rm W}(c\,
x^2_{\rm Q})\right]^2} +\left(\frac{0.77
g}{8\sqrt{2}\pi}\right)^2 +
\left(\frac{9\pi}{4}\right)^2\right\}^{1/2}~,
\end{equation}
to obtain $\xi$, and then the CS contributions for a given value of $g$.
Our results are summarised in Fig.~\ref{PRL4}

Within this approach for D-term inflation, our
findings are different than within the framework of SUSY~\cite{rs},
unless $\lambda \gtrsim 10^{-3}$ or $g \lesssim 10^{-4}$; in these
cases we are in the SUSY limit. The CS contribution to
the CMB turns out to be model-dependent, with however the robust
results that CS contribution is not constant, nor is it always
dominant, in contradiction to Ref.~\cite{rj}.
This implies that contrary to what is often assumed,
D-term inflation is still an open possibility. Our analysis~\cite{rs}
shows that $g\gtrsim 1$ necessitates
multiple-stage inflation, since otherwise we cannot have $N_{\rm
Q}=60$, while $g\gtrsim 2\times 10^{-2}$ is incompatible with the
allowed CS contribution to the WMAP measurements.  For $g\lesssim 2\times
10^{-2}$, we can also constrain the superpotential coupling $\lambda$
and get $\lambda \lesssim 3\times 10^{-5}$. This limit was already
found in the SUSY framework~\cite{rs} and it is in agreement with the finding
$\lambda\lesssim {\cal O}(10^{-4}-10^{-5})$ of
Ref.~\cite{jap}. However, we disagree with the result of
Ref.~\cite{jap} stating that $\lambda$ can be of order 1 using the curvaton
mechanism, since this is forbidden by the gravitino
constraint~\cite{rs}. This upper bound on $\lambda$ is also in
agreement with the one found in Ref.~\cite{kl} in the context of
P-term inflation.
 SUGRA corrections impose~\cite{rs} in addition
a lower limit to the coupling $\lambda$. If for example $g=10^{-2}$,
the cosmic strings contribution imposes  $10^{-8}\lesssim \lambda
\lesssim 3\times 10^{-5}$.  This constraint induced by the CMB
measurements is expressed as a single contraint on the Fayet-Iliopoulos
term $\xi$, namely $\sqrt\xi \lesssim 2.3\times 10^{15}~{\rm GeV}$.

We would like to bring to the attention of the reader that in the
above study we have neglected the quantum gravitational effects, which
would lead to a contribution to the effective potential, even though
$S_{\rm Q}\sim {\cal O}(10 M_{\rm Pl})$.  Our analysis is however
still valid, since the effective potential given in
Eq.~(\ref{VexactDsugra}) satisfies the conditions~\cite{lindebook}
$V(|S|)\ll M_{\rm Pl}^4$ and $m^2_S={\rm d}^2 V/{\rm d}S^2 \ll M_{\rm
  Pl}^2$, and thus the quantum gravitational corrections $[\Delta
V(|S|)]_{\rm QG}$ are negligible when compared to the effective
potential $V_{\rm eff}^{\rm D-SUGRA}$.

In conclusion, our study has shown that cosmic strings are allowed
within SUSY GUTs and they may have an important cosmological
r\^ole. We proved that CS formed at the GUT scale are consistent with the
recent CMB measurements. The implications of our findings are
important.  High energy physics tells us that cosmic strings are
expected to be formed at the end of hybrid inflation~\cite{rjs}. These objects
may lead to important cosmological consequences. Measurements from the
realm of cosmology on the other hand constrain the possible r\^ole of
cosmic strings, leading to constraints on the free parameters of the
employed SUSY/SUGRA models. Thus, cosmology gives us, in return,
information about high energy physics. This is indeed the beauty of
this subject.

\section*{Acknowledgements}
It is a pleasure to thank P.\ Brax, N.\ Chatillon,
 G. Esposito-Far\`ese, J.\ Garcia-Bellido, R.\ Jeannerot, A.\ Linde,
 J.\ Martin, P.\ Peter,
and C.\ -M.\  Viallet for discussions and comments.

\end{document}